\documentclass[%
 aip,
 amsmath,amssymb,
 reprint,%
]{revtex4-1}

\usepackage{graphicx}% Include figure files
\usepackage{dcolumn}% Align table columns on decimal point
\usepackage{bm}% bold math
\usepackage{textcomp}
\usepackage{gensymb}
\usepackage[utf8]{inputenc}
\usepackage[T1]{fontenc}
\usepackage{mathptmx}
\usepackage{etoolbox}
\usepackage{siunitx}
\usepackage{textgreek}
\usepackage{xr}
\usepackage{bm}
\usepackage{xcolor}
\usepackage{tikz}
\newcommand{\et}{\textit{et al.~}}
\newcommand{\revise}[1]{\textcolor{black}{#1}}
\newlength{\wordlen}

\begin{document}

\preprint{APS/123-QED}

\title{Transport of skyrmions by surface acoustic waves}% Force line breaks with \\
\thanks{Authors to whom correspondence should be addressed: John E Cunningham, J.E.Cunningham@leeds.ac.uk; Thomas A Moore, T.A.Moore@leeds.ac.uk. Present address of Jintao Shuai: National Creative Research Initiative Center for Spin Dynamics and Spin-Wave Devices, Nanospinics Laboratory, Research Institute of Advanced Materials, Department of Materials Science and Engineering, Seoul National University, Seoul 08826, Republic of Korea}%

\author{Jintao Shuai}
\affiliation{School of Physics and Astronomy, University of Leeds, Woodhouse Lane, Leeds, LS2 9JT, UK}

\author{Luis Lopez-Diaz}
\affiliation{Department of Applied Physics, Universidad de Salamanca, Salamanca, Spain}

\author{John E. Cunningham}
\affiliation{School of Electronic and Electrical Engineering, University of Leeds, Woodhouse Lane, Leeds, LS2 9JT, UK}

\author{Thomas A. Moore}
\affiliation{School of Physics and Astronomy, University of Leeds, Woodhouse Lane, Leeds, LS2 9JT, UK}

\date{\today}% It is always \today, today,

\begin{abstract}
Magnetic skyrmions in thin films with perpendicular magnetic anisotropy (PMA) are promising candidates for magnetic memory and logic devices, making the development of ways to transport skyrmions efficiently in a desired trajectory of significant interest. Here, we investigate the transport of skyrmions by surface acoustic waves (SAWs) via several modalities using micromagnetic simulations. We show skyrmion pinning sites created by standing SAWs at anti-nodes and skyrmion Hall-like motion without pinning driven by travelling SAWs. We also show how orthogonal SAWs formed by combining a longitudinal travelling SAW and a transverse standing SAW can be used for the 2D positioning of skyrmions. Our results also suggest SAWs offer a viable approach to the transport of multiple skyrmions along a multichannel racetrack.
\end{abstract}

\maketitle
% \section{Introduction}
Magnetic skyrmions, which are topologically protected particle-like magnetic structures, show significant potential in applications including data storage and processing devices~\cite{nagaosa2013topological,fert2013skyrmions}.
Skyrmions in thin films can be manipulated by spin-polarised current via spin-transfer torque or spin-orbit torque owing to the large spin-orbit coupling in adjacent heavy metals~\cite{torrejon2014interface,litzius2020role,woo2016observation,caretta2018fast,woo2018current}.
However, these methods require a high current density, which can cause Joule heating thereby wasting energy and affecting the stability of skyrmions. In addition, skyrmions typically show both longitudinal and transverse motion owing to the skyrmion Hall effect, which can cause the annihilation of skyrmions at device edges, thus complicating device realisations~\cite{nagaosa2013topological,juge2019current,woo2018current}.\par
Typically, the stability of skyrmions in thin films is a result of the balance between the perpendicular magnetic anisotropy (PMA) and the interfacial Dzyaloshinskii-Moriya interaction (DMI) induced by the broken interfacial inversion symmetry~\cite{bhattacharya2020creation,dzyaloshinsky1958thermodynamic,moriya1960anisotropic,barker2023breathing}. To avoid Joule heating, one can modify the anisotropy of thin films using strain to control magnetisation~\cite{shepley2015modification,matsukura2015control,radaelli2014electric,shirahata2015electric}.
For instance, Wang \et created skyrmions in Pt/Co/Ta multilayer nano-dot systems using an electrical field-induced strain\cite{wang2020electric}, while Ba \et demonstrated the creation, reversible deformation, and annihilation of skyrmions in a Pt/Co/Ta multilayer thin film using strain by applying an electric field to a \revise{PMN-PT (lead magnesium niobate-lead titanate)} substrate~\cite{ba2021electric}.\par

The dynamic strain induced by surface acoustic waves (SAWs) has also been suggested as an attractive approach to control thin film magnetisation~\cite{shuai2022local,dean2015sound,thevenard2016strong,adhikari2021surface,edrington2018saw,ryu2013chiral,cao2021surface,li2014acoustically,shuai2023separation,shuai2023surface,miyazaki2023trapping,yang2021acoustic,yang2023magnetic}.
In particular, Yokouchi \et experimentally observed the creation of skyrmions by SAWs in a Pt/Co/Ir thin film owing to the inhomogeneous effective torque arising from both SAWs and thermal fluctuations via magnetoelastic coupling~\cite{yokouchi2020creation}.
Nepal \et theoretically studied the dynamical pinning of skyrmion bubbles at the anti-nodes of standing SAWs in a FePt nano-wire, revealing the strain gradient induced by SAWs is the driving force for skyrmion motion~\cite{nepal2018magnetic}.
Chen \et demonstrate both experimentally and micromagnetically that the longitudinal leaky SAW provide both a strain and a thermal effect that can be used to create skyrmions~\cite{chen2023ordered}. An electrical current (with the direction perpendicular to the propagation direction of the SAWs) together with the standing SAWs was then applied to move the skyrmions. In the presence of the SAWs, the transverse motion of the skyrmion was suppressed and the skyrmion showed motion in a straight line. 
Skyrmion motion driven by SAWs has also been experimentally demonstrated by Yang \et~\cite{yang2024acoustic}. This study paves the way for the experimental feasibility of neuromorphic computing using SAW-controlled synapse devices based on skyrmions, for instance, as highlighted in Chen \et study~\cite{chen2021surface}.
Another group presents the skyrmion moved in a straight line theoretically with an analytical model and micromagnetic simulations~\cite{chen2023suppression}. The electrical current provides driving force for skyrmion motion in both horizontal and transverse directions. The standing SAWs, on the other hand, create pinning channels suppressing the transverse motion of the skyrmion.
SAW control of magnetism has a number of potential advantages. SAWs can be generated by voltage instead of current making it attractive from the energy-efficiency perspective, while them can propagate over distances of several millimetres with very little power loss, and can also allows one pair of electrodes to control multiple devices. In addition, pinning sites can be created by electrodes remotely, which potentially allows one to control skyrmions without complex design.\par
\revise{Despite the advantages of SAW-driven skyrmion motion, there are still knowledge gaps in current research that need to be addressed. For example, the potential for suppressing SAW-induced skyrmion Hall-like motion using orthogonal SAWs -- formed by combinations of travelling and standing waves -- has not yet been demonstrated, nor have the dynamics of skyrmions in such SAW configurations been investigated.}
In this work, we demonstrate the transport of skyrmions by several different modalities of SAWs using micromagnetic simulations, and propose a multi-channel skyrmion racetrack that can transport multiple skyrmions in parallel. 
We first of all show that standing SAWs are able to introduce skyrmion motion from a node to the nearest anti-node, where the skyrmion is pinned as reported in~\cite{nepal2018magnetic}. We then demonstrate travelling SAW-induced skyrmion continuous motion with Hall-like motion. To confine the skyrmion motion in a straight line, the orthogonal SAWs, which are created by travelling SAWs propagating in one direction and standing SAWs in the transverse direction, are then applied.

% \section{Computational details}
\begin{figure*}[ht!]
\includegraphics[width=1\textwidth]{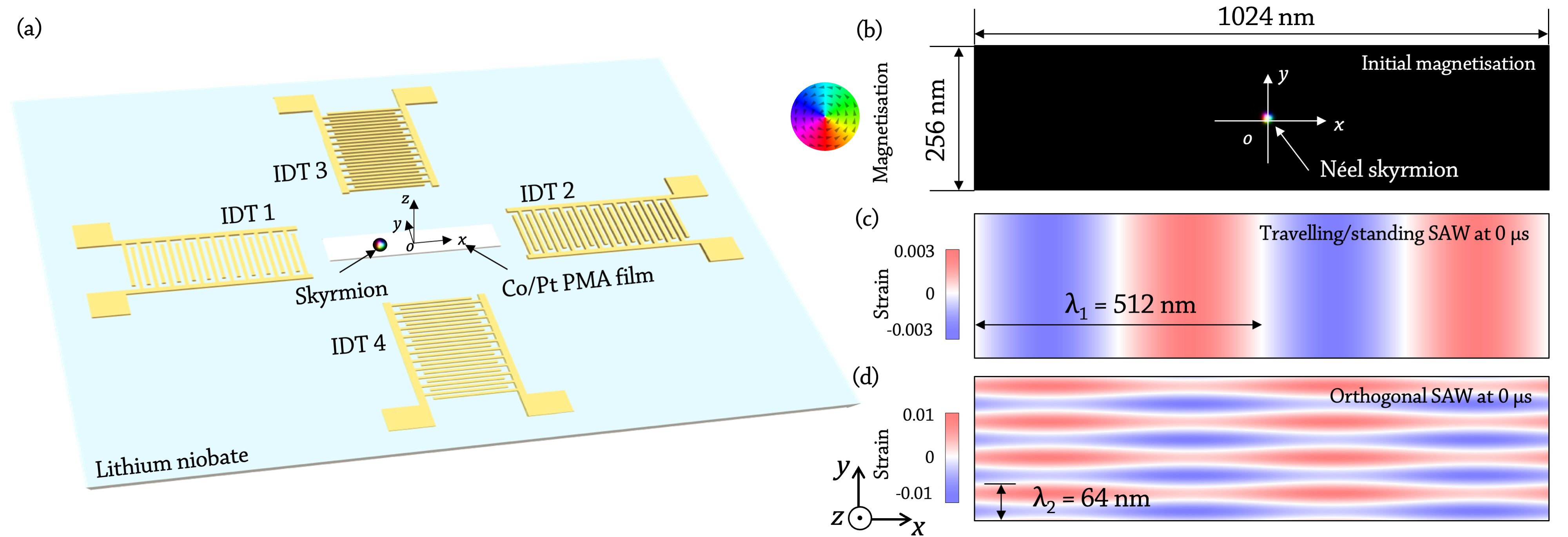}
\caption{\label{fig:config}
(a) Diagram of the proposed device (not to scale). A skyrmion is initialised in a Co/Pt PMA thin film grown onto a lithium niobate substrate. The thin film is surrounded by two pairs of \revise{interdigitated transducers (IDTs, labelled as IDT1 to IDT4)}, which can be used to generate SAWs in different modalities. (b) An example of the initial magnetisation. A N\'eel skyrmion is initialised and relaxed in thin film with computational dimensions of $1024\times256\times1$ nm\textsuperscript{3}. (c) Strain spatial profile of travelling SAWs and standing SAWs at 0 \textmu s. Wavelength and strain amplitude of SAWs are 512 nm and 0.003, respectively. (d) Spatial strain profile of orthogonal SAWs (longitudinal travelling SAWs and transverse standing SAWs) at 0 \textmu s. Wavelengths and strain amplitudes of standing SAWs and travelling SAWs, which together form orthogonal SAWs, are 512 nm and 0.003, and 64 nm and 0.007, respectively. The maximum strain is 0.01. \revise{The color code for the in-plane magnetisation component is shown by a color wheel. The Cartesian coordinates define the spatial orientation of the thin film. }}
\end{figure*}

Fig.~\ref{fig:config}a shows the schematic diagram of the proposed device. A Co/Pt PMA thin film (with dimensions $1024\times256\times1$ nm\textsuperscript{3}) is coupled to a 128\degree\ Y-cut lithium niobate substrate and surrounded by two pairs of \revise{interdigitated transducers (IDT, labelled as IDT1 to IDT4)}. \revise{We consider scenarios where the size of the skyrmions is significantly smaller than the width of the thin film. Therefore, periodic boundary conditions are employed along the $y$ axis. The edge effects on skyrmion motion are thus neglected.} As shown in Fig.~\ref{fig:config}a and b, the origin of the thin film geometry is located at its centre, with $x$ and $y$ correspond to the longitudinal and transverse directions in the thin film, respectively. The electrode spacing of the first (IDT1 and IDT2) and second (IDT3 and IDT4) pair of IDTs was 128 nm and 16 nm, respectively, producing SAWs with wavelength $\lambda_1$ (see Fig.~\ref{fig:config}c) and $\lambda_2$ (see Fig.~\ref{fig:config}d) of 512 nm and 64 nm, respectively. SAWs with different propagation modes can be achieved by applying rf signals to one or more IDTs: travelling SAWs propagating in $x$ direction can be generated by applying rf signals to IDT1; standing SAWs in the $x$ direction can be formed by applying rf signals to IDT1 and IDT2 simultaneously; while orthogonal SAWs consisting of horizontal travelling SAWs and transverse standing SAWs, can be formed by applying rf signals to IDT1, IDT3 and IDT4 at the same time. N\'eel skyrmions were initialised and relaxed at different positions within the perpendicular magnetised thin film in order to study the SAW effect on their motion (Fig.~\ref{fig:config}b). The strain amplitude of standing SAWs and travelling SAWs was 0.003. To generate orthogonal SAWs, amplitudes of the longitudinal travelling SAWs and transverse standing SAWs were 0.003 and 0.007, respectively. Fig.~\ref{fig:config}c and d show the spatial strain profile of travelling SAWs, standing SAWs and orthogonal SAW at 0 \textmu s, respectively.\par
Micromagnetic simulations were performed using Mumax3~\cite{Vansteenkiste2014,Exl2014,Mulkers2017}, based on the Landau-Lifshitz-Gilbert (LLG) equation
\begin{equation}
    \frac{{d\mathbf{M}}}{{dt}} = -\gamma \mathbf{M} \times \mathbf{H}_\text{eff} + \frac{\alpha}{M_\mathrm{s}}\mathbf{M}  \times \frac{{d\mathbf{M}}}{{dt}}\text{,}
\end{equation}
where $\mathbf{M}$ is the magnetisation vector, $M_\mathrm{s}$ is the saturation magnetisation, $t$ is the time, $\gamma$ is the gyromagnetic ratio, $\mathbf{H}_\text{eff}$ is the effective magnetic field acting on the magnetisation, and $\alpha$ is the Gilbert damping constant. 
% In our simulations, the total free energy density $f_{total}$ is:
% \begin{equation}
%     f_{total}=f_{ani}+f_{stray}+f_{ex}+f_{DMI}+f_{mel}
% \end{equation}
% where $f_{ani}$ is the magneto-crystalline uniaxial anisotropy energy density, $f_{stray}$ is the magnetostatic stray field energy density, $f_{ex}$ is the magnetic exchange energy density, $f_{DMI}$ is the interfacial DMI energy density, and 
The magneto-elastic energy density, $E_\mathrm{me}$, can be expressed as
\begin{equation}
     E_\mathrm{me} = B_1\sum_{i = x, y, z}m_i^2\epsilon_{ii} + B_2\sum_{i \neq j}m_im_j\epsilon_{ij}
     \label{eqn:E_me}
\end{equation}
where $B_1$ and $B_2$ are the magneto-elastic coefficients, and $\epsilon_{ij}$ is the strain tensor.
The effective field can be described as the first derivative of the free energy with respect to magnetisation. The effective field introduced by the magnetoelastic interaction can thus be expressed as
\begin{equation}
    \mathbf{H}^{(i)}_\mathrm{me} = -\frac{2}{\mu_0 M_\mathrm{s}} \left(B_{1} m_{i} \epsilon_{ii} + B_{2} \sum_{j: j \neq i} m_{j} \epsilon_{ij} \right),
    \label{eqn:H_me}
\end{equation}
where $\mathbf{H}^{(i)}_\mathrm{me}$ is the component of the effective field along the axis labeled by $i$.
The material parameters were set corresponding to a Co/Pt thin film as follows~\cite{sampaio2013nucleation,gutjahr2000magnetoelastic}: 
saturation magnetisation $M_\mathrm{s} = \SI{5.8e5}{A/m}$, exchange constant $A_\mathrm{ex} = \SI{1.5e-11}{J/m}$, 
Gilbert damping coefficient $\alpha$ is 0.1,
perpendicular magnetic anisotropy $K_\mathrm{u}$ is $\SI{8.0e5}{J/m^{3}}$,
an interfacial DMI strength of $\SI{3.0e-3}{J/m^{2}}$,
and magneto-elastic coupling $B_1$ of $\SI{2.0e7}{J/m^{3}}$.

We note the film has been set to be acoustically thin - that is sufficiently thin and rigid compare to the SAW wavelength that only the in-plane strain component ($\epsilon_{ii}$, $i=x,y$) needs to be taken into consideration~\cite{campbell2012surface,yokouchi2020creation,dean2015sound,nepal2018magnetic}. Thus, $B_2$ in Equation~\ref{eqn:E_me} and~\ref{eqn:H_me} was set to 0.
The in-plane strain of standing SAWs along $x$-axis (for standing SAW case) and $y$-axis (for orthogonal SAW case), $\epsilon_{xx}^\mathrm{S}$ and $\epsilon_{yy}^\mathrm{S}$, were implemented as
\begin{equation}
    \epsilon_{xx}^\mathrm{S} = A_{xx}^\mathrm{S} \sin(k_{x}^\mathrm{S}x)\cos(\omega_{x}^\mathrm{S} t),
    \label{eqn:standing_SAW_x}
\end{equation}
\begin{equation}
    \epsilon_{yy}^\mathrm{S} = A_{yy}^\mathrm{S} \sin(k_{y}^\mathrm{S}y)\cos(\omega_{y}^\mathrm{S} t),
    \label{eqn:standing_SAW_y}
\end{equation}
respectively, where $k_{x}^\mathrm{S}$, $k_{y}^\mathrm{S}$ and $\omega_{x}^\mathrm{S}$, $\omega_{y}^\mathrm{S}$ are the wavenumber and angular frequency of the standing SAW in $x$ and $y$ directions, respectively, $A_{xx}^\mathrm{S}$ and $A_{yy}^\mathrm{S}$ are the corresponding amplitudes of the strain, and $t$ is time.

The in-plane strain of the travelling SAW along the $x$ direction, $\epsilon_{xx}^\mathrm{T}$, was implemented as
\begin{equation}
    \epsilon_{xx}^\mathrm{T} = A_{xx}^\mathrm{T} \sin(\omega_{x}^\mathrm{T} t - k_{x}^\mathrm{T} x),
    \label{eqn:travelling_SAW_x}
\end{equation}
where $k_{x}^\mathrm{T}$, $\omega_{x}^\mathrm{T}$, and $A_{xx}^\mathrm{T}$ are the wavenumber, angular frequency, and amplitude of the travelling SAW, respectively, and $t$ is time. The velocity of SAWs propagating in lithium niobate is 4000 m/s~\cite{paskauskas1995velocity}.
% \begin{equation}
%     \epsilon_{xx}^\mathrm{T} = A_{xx}^\mathrm{T} (\sin(\omega_{x}^\mathrm{T} t)\cos(k_{x}^\mathrm{T} x)-\cos(\omega_{x}^\mathrm{T} t)\sin(k_{x}^\mathrm{T} x) )
% \end{equation}

% \section{Results and discussions}
\begin{figure*}
\includegraphics[width=1\textwidth]{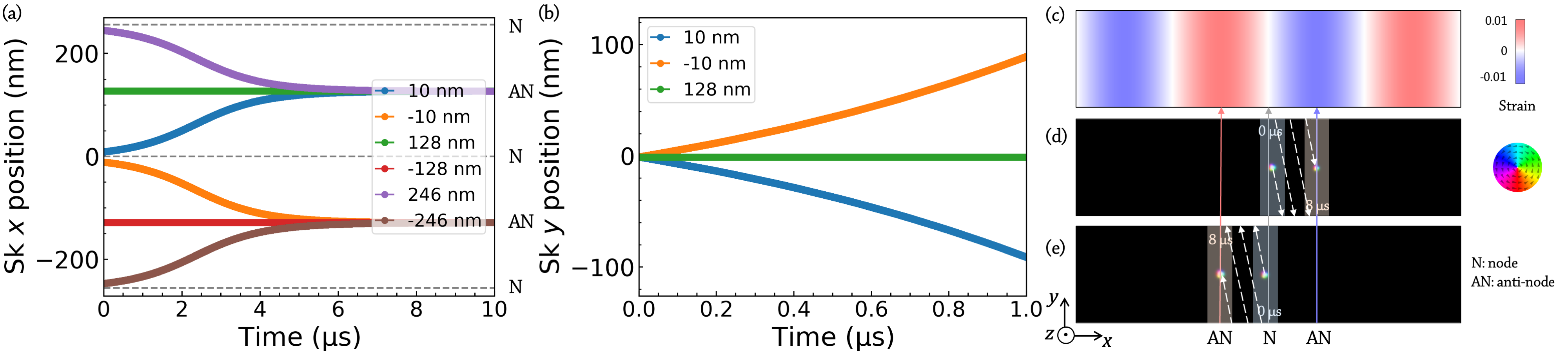}
\caption{\label{fig:standingSAW} 
Translation of \revise{skyrmion (sk)} in $x$ (a) and $y$ (b) directions driven by standing SAWs. Numbers in the legend indicate initial $x$ positions of skyrmions. (c) Spatial strain profile of standing SAWs at 0 \textmu s. Examples of the skyrmion initial and final positions driven by standing SAWs: skyrmion initialised at -10 nm (d) and 10 nm (e) along $x$ direction. \revise{The shaded areas in (d) and (e) indicate the initial and final position of the skyrmions at 0 \micro s and 8 \micro s, respectively. The dashed line is the trajectory of the skyrmions with periodic boundary conditions.} N and AN demote the node and anti-node, respectively. \revise{The color code for the in-plane magnetisation component is shown by a color wheel. The Cartesian coordinates define the spatial orientation of the thin film.}}
\end{figure*}
The dynamics of rigid magnetic textures subject to an external driving force can be described by the Thiele equation. SAW-driven skyrmion motion at position $\mathbf{R}(X, Y)$ can be expressed as:
\begin{equation}
    \mathbf{G}\times \mathbf{v} - \alpha \mathbf{D} \cdot \mathbf{v} + \mathbf{F}(\mathbf{R}) = 0,
    \label{eqn:thiele_equation}
\end{equation}
where $\mathbf{G} = (0, 0, -4 \pi Q)$ is the gyromagnetic coupling vector with topological charge $ Q = \frac{1}{4 \pi}\int\mathbf{m} \cdot (\frac{\partial \mathbf{m}}{\partial x} \times \frac{\partial \mathbf{m}}{\partial y})dxdy$, 
$\mathbf{v} = \mathbf{\dot{R}}$ is the velocity of the skyrmion,
$\mathbf{D}$ is a dissipative tensor in the form of $\bigl(\begin{smallmatrix} D & 0 \\ 0 & D \end{smallmatrix} \bigr)$ for an isolated skyrmion with $D = \int \frac{\partial \mathbf{m}}{\partial x} \cdot \frac{\partial \mathbf{m}}{\partial y}dxdy$, 
and $\mathbf{F}(\mathbf{R})$ is the effective force per unit thickness due to the strain induced by SAWs.
The effective force $\mathbf{F}(\mathbf{R})$ generated by the strain $\epsilon_{ii}$ ($i=x,y$) of the SAWs at the given time $t$ can be expressed as~\cite{nepal2018magnetic,miyazaki2023trapping}:
\begin{equation}
    \mathbf{F}(\mathbf{R},t) = \frac{1}{k_i^2} C_{ii} \nabla \epsilon_{ii}(\mathbf{r},t)|_{\mathbf{r}=\mathbf{R}(t)},
\end{equation}
where $C_{ii}$ is a shape factor due to strain and their values depend upon the shape and size of the bubble~\cite{nepal2018magnetic,miyazaki2023trapping}, and $\mathbf{r}$ is the centre of the skyrmion at the give time $t$. 
The skyrmion velocity is proportional to the strain gradient.
The SAW-induced force acting upon the skyrmion is only in the direction of the applied SAWs, whereas the force exerted by the SAWs in the direction perpendicular to their application is zero, as there is no strain gradient present in that direction. 
The first term in Equation~\ref{eqn:thiele_equation} represents the Magnus force and moves the skyrmion in transverse direction. 
The Magnus force thus is normal to the direction of the skyrmion velocity. 
The second term in Equation~\ref{eqn:thiele_equation} is the dissipative force due to damping.
There are therefore in total three forces acting upon a skyrmion when driven by one set of SAWs (standing or travelling SAWs) alone, i.e., the driving force provided by SAWs in the direction of the applied SAWs ($\mathbf{F}_\mathrm{S}$ for standing SAWs or $\mathbf{F}_\mathrm{T}$ for travelling SAWs), Magnus force $\mathbf{F}_\mathrm{M}$ perpendicular to the direction of skyrmion velocity, and dissipative force $\mathbf{F}_\mathrm{Dis}$ due to damping.\par

We first study the skyrmion motion driven by standing SAWs alone, and in particular the effect of nodes and anti-nodes of standing SAWs on skyrmion motion. The standing SAWs were applied along the $x$ direction using Equation~\ref{eqn:standing_SAW_x}.
The dynamic strain waves form a strain gradient, which periodically changes between nodes and anti-nodes, providing a driving force for skyrmion motion in the $x$ direction.
However, this strain gradient $\nabla\epsilon_{xx}$ vanishes at the anti-nodes of standing SAWs. 
The induced strain gradient pushes the skyrmion moving towards the anti-nodes, with pinning therefore occurring at the anti-nodes. 
The wavelength of the standing SAWs in this case is 512 nm, meaning the nodes are located at $X=$ --256, 0, and 256 nm, and the anti-nodes are located at $X=$ --128 and 128 nm.
The amplitude of the SAWs is 0.003.
Skyrmions were initialised at a few different positions along the $x$-axis, with some near the nodes ($X=$ --246, --10, 10, and 246 nm), and some at the anti-nodes $X=$ --128 and 128 nm.
It is shown in Fig.~\ref{fig:standingSAW}a and b that the skyrmions located at the anti-node are pinned at the position without any motion either $x$ or $y$ directions due to the absence of the strain gradient at the anti-nodes.
On the other hand, when the skyrmions are initialised at the nodes, they move towards the closest anti-nodes where they are pinned afterwards (see Fig.~\ref{fig:standingSAW}a). For instance, the skyrmions located at $X=$ --10 and 10 nm move towards $X=$ --128 and 128 nm, respectively. 
Meanwhile, these skyrmions also show transverse motion due to the Magnus force as expected.
Fig.~\ref{fig:standingSAW} d and e show examples of skyrmion motion from initial positions (10 nm and -10 nm) to final positions at 8 \textmu s.
The skyrmion velocity is proportional to the strain gradient, which decreases from the maximum value at the node to zero at the anti-node. The skyrmion velocity is thus not constant, and decreases when the skyrmion is getting closer to the anti-node. These results are in good agreement with Nepal \et~\cite{nepal2018magnetic}. \par

\begin{figure}
\includegraphics[width=0.45\textwidth]{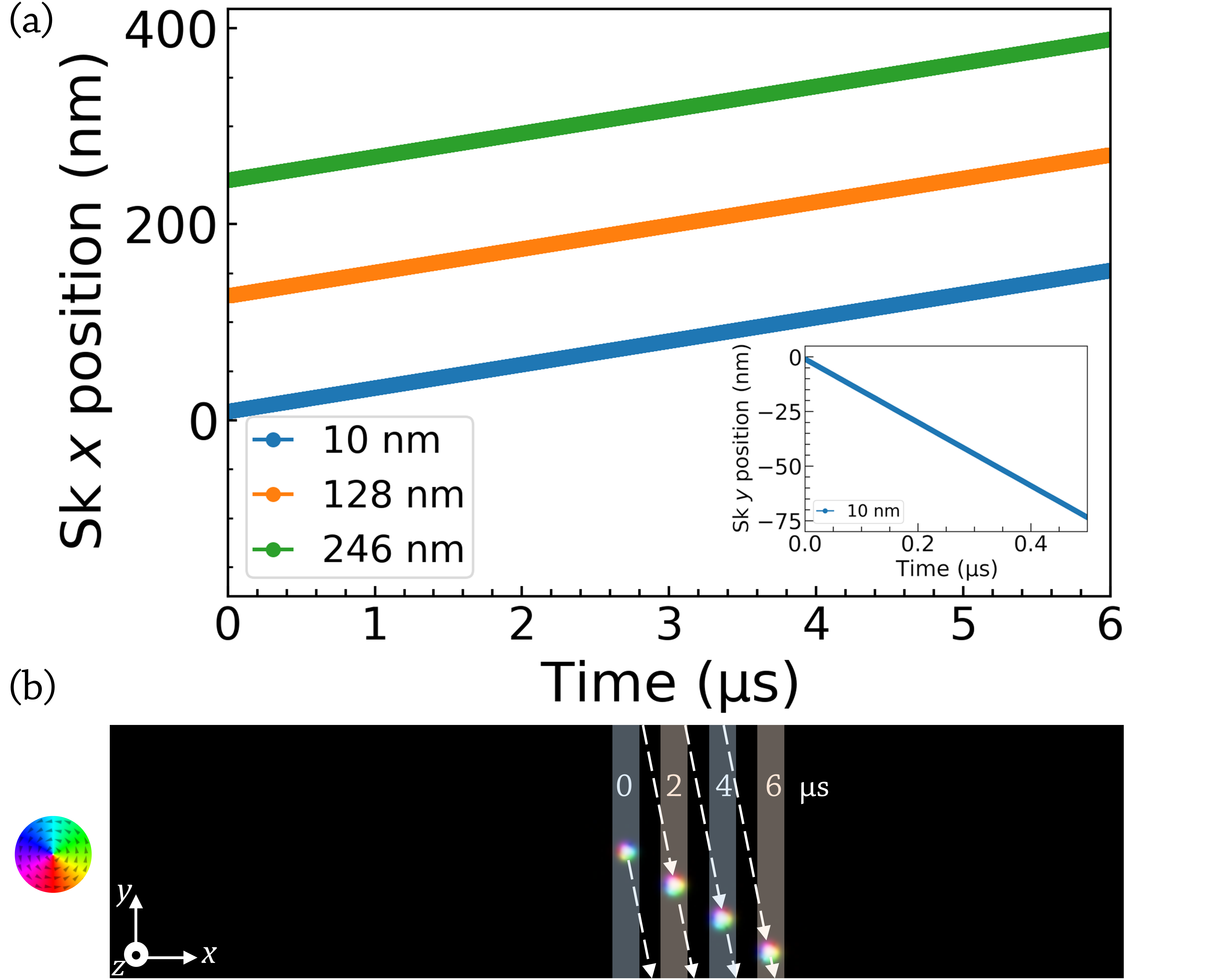}
\caption{\label{fig:travellingSAW} 
(a) Translation of \revise{skyrmion (sk)} in the $x$ direction driven by travelling SAWs. Numbers in the legend indicate the initial positions of skyrmions along the track. The insert graph shows the skyrmion motion in the $y$ direction during the first 0.5 \textmu s. (b) Snapshots of skyrmion motion from 0 to 6.0 \textmu s. \revise{The shaded areas indicate the skyrmion positions at the corresponding time labelled on the shaded area from 0 to 6 \micro s, respectively. The dashed line is the trajectory of the skyrmions with periodic boundary conditions.} \revise{The color code for the in-plane magnetisation component is shown by a color wheel. The Cartesian coordinates define the spatial orientation of the thin film.}}
\end{figure}

Secondly, we study the skyrmion motion driven by travelling SAWs. 
In the presence of the travelling SAWs, unlike the standing SAWs, the nodes and anti-nodes of the travelling SAWs are not stationary in space. 
The nodes and anti-nodes of the travelling SAWs propagate along the applied direction as does the strain gradient ($\nabla \epsilon_{xx}$). This means that the skyrmions experience a strain gradient along the SAW directions at any given location. Therefore, it is expected for skyrmions to move continuously without pinning.
To compare the difference of the travelling SAW- and standing SAW-driven skyrmion, we set the wavelength and amplitude of the travelling SAW as 512 nm and 0.003, respectively, which are the same values as those of the standing SAW discussed above.
Fig.~\ref{fig:travellingSAW}a shows the translation of skyrmions in the $x$ direction driven by travelling SAWs. 
Unlike the skyrmion motion driven by standing SAWs, regardless of initial positions ($X=$ 10, 128, and 246 nm), the skyrmions move continuously in the $x$ direction at a constant speed $v_x$ of 2.40 cm/s without any pinning with the application of travelling SAWs (Fig.~\ref{fig:travellingSAW}a). This is because skyrmions experience nodes and anti-nodes at all positions with the application of travelling SAWs, which provide a dynamic but continuous strain gradient (driving force). Skyrmions also exhibit Hall-like motion in the $y$ direction due to the Magnus force (see the insert graph in Fig.~\ref{fig:travellingSAW}a). The skyrmion speed in the $y$ direction $v_y$ is 15.10 cm/s. Fig.~\ref{fig:travellingSAW}b shows an example of skyrmion position changes with time driven by travelling SAWs. \par

\begin{figure}
\centering
\includegraphics[width=0.45\textwidth]{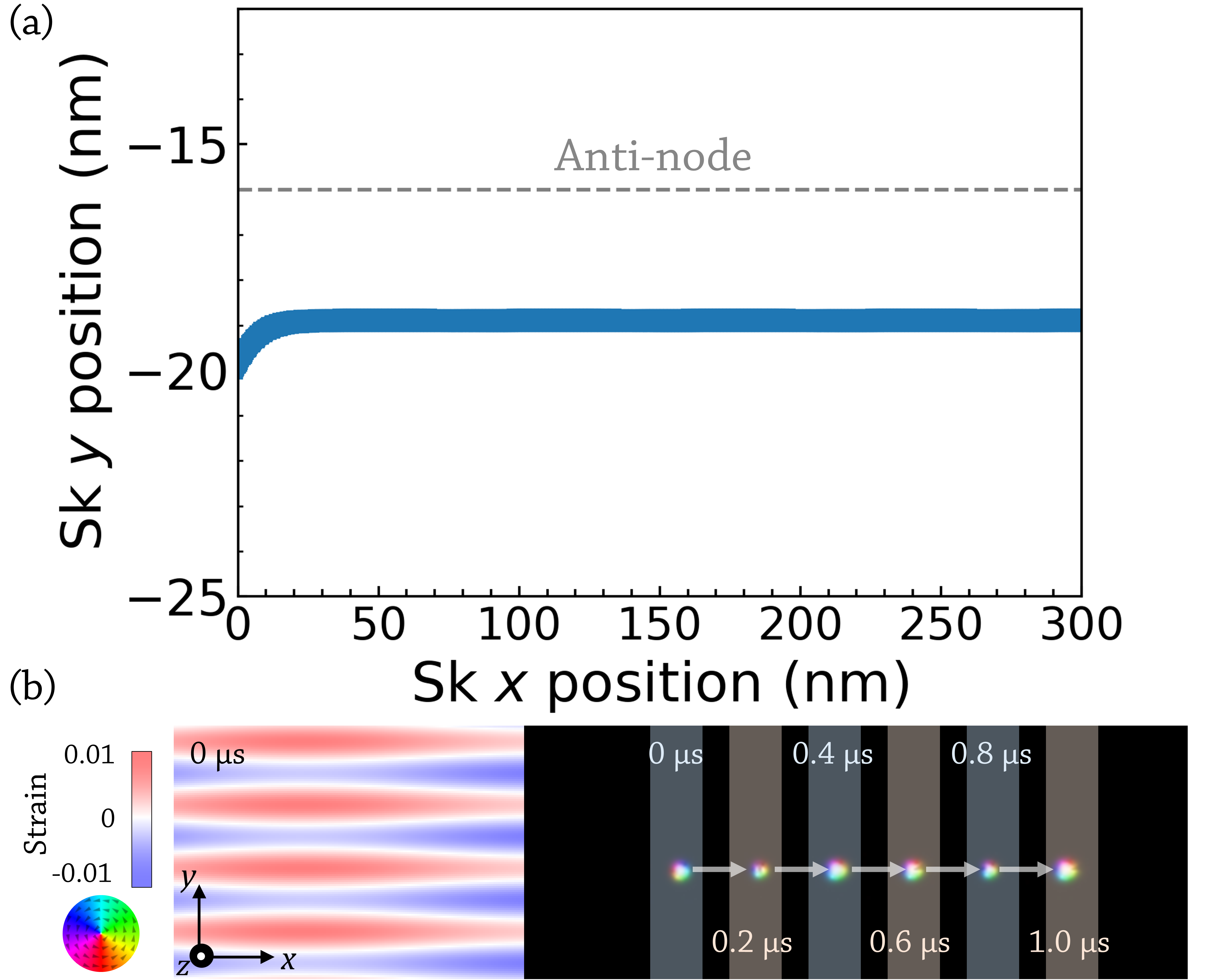}
\caption{\label{fig:orthogonalSAW} 
(a) Translation of \revise{skyrmion (sk)} in $x$ and $y$ directions driven by orthogonal SAWs. The insert graph shows the skyrmion motion in the $y$ direction during the first 0.1 \textmu s. (b) Left-hand side of the graph shows the spatial stain profile of orthogonal SAWs at 0 \textmu s. Right-hand side of the graph shows the snapshots of the skyrmion motion from 0 to 1.0 \textmu s. \revise{The shaded areas indicate the skyrmion positions at the corresponding time labelled on the shaded area from 0 to 1 \micro s, respectively.} The arrows indicate the direction of the skyrmion motion, which is in line with the anti-node of standing SAWs in orthogonal SAWs. \revise{The color code for the in-plane magnetisation component is shown by a color wheel. The Cartesian coordinates define the spatial orientation of the thin film.} }
\end{figure}

\begin{figure*}
\includegraphics[width=1\textwidth]{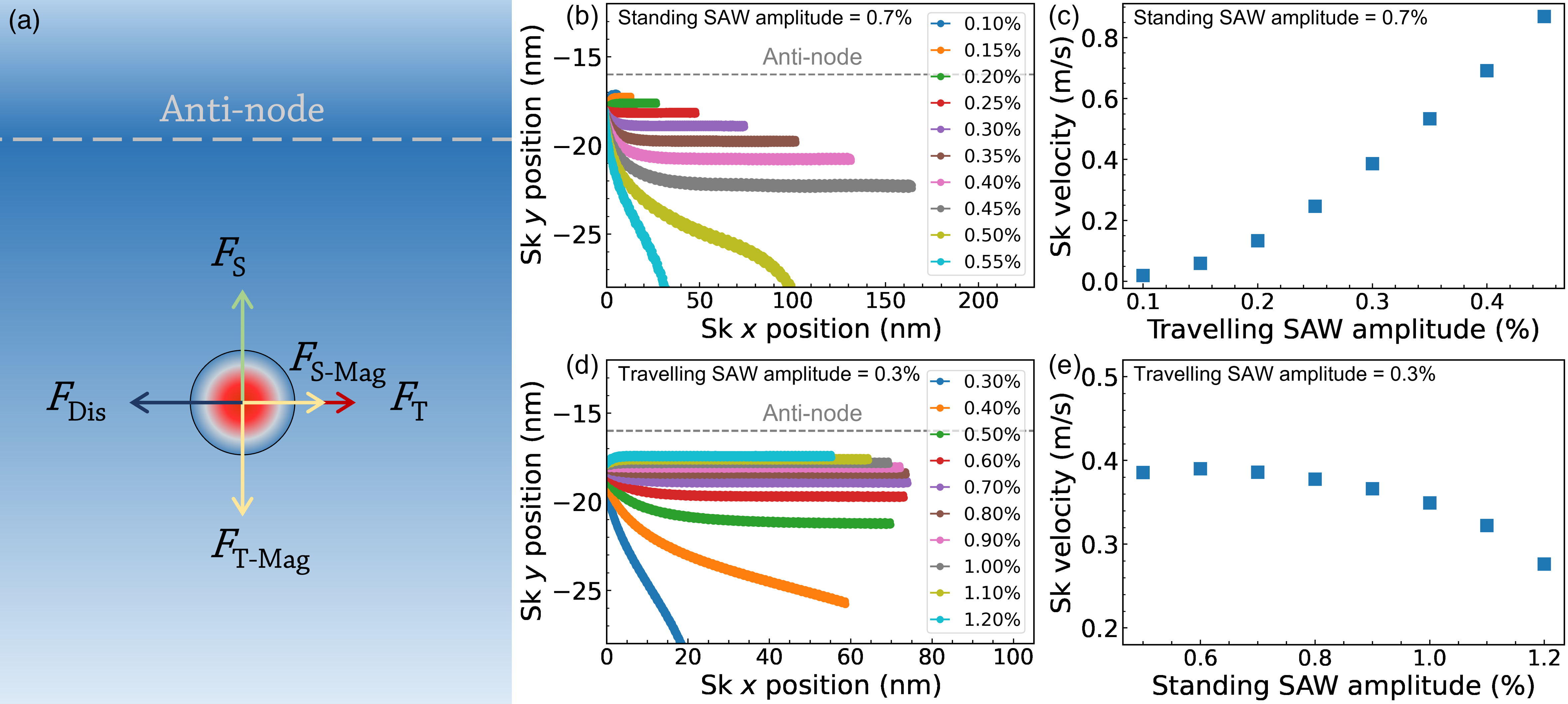}
\caption{\label{fig:sk_velocity} 
(a) Schematics of the different forces acting on the skyrmion. \revise{Skyrmion (sk)} trajectory (b) and velocity (c) for a fixed amplitude of the standing SAW and different amplitudes of the traveling SAW.  Skyrmion trajectory (b) and velocity (c) for a fixed amplitude of the traveling SAW and different amplitudes of the standing SAW.}
\end{figure*}

As discussed above, the skyrmion transverse motion is induced by the the Magnus force. To move skyrmion continuously in a straight line, it is necessary to suppress this Magnus force. We propose moving the skyrmion in the $x$ direction using orthogonal SAWs formed by travelling SAWs propagating in the $x$ direction and standing SAWs applied along the $y$ direction. With this configuration, we expect to move skyrmions continuously in the $x$ direction with the continuous driving force provided by travelling SAWs and to confine the motion in $y$ direction by the pinning ``channels'' using the anti-nodes of the standing SAWs along the $y$ direction. In this simulation, the wavelengths and strain amplitudes of travelling SAWs and standing SAWs were 512 nm and 0.003, and 64 nm and 0.007, respectively (see Fig.~\ref{fig:config}d). Fig.~\ref{fig:orthogonalSAW}a shows the translation of a skyrmion. The skyrmion moves continuously in the $x$ direction with a very limited motion of $\sim$1 nm in the $y$ direction at the beginning (see insert graph in Fig.~\ref{fig:orthogonalSAW}a), when the skyrmion tries to move from its initial position towards the pinning channel that standing SAWs create. Note that the skyrmion velocity significantly increases to 38.64 cm/s compared to that of travelling SAW-induced skyrmion motion (2.40 cm/s). This is because the another Magnus force along the x-axis is formed when the transverse standing SAWs are applied to the skyrmion ($F_\mathrm{S-Mag}$ in Fig. \ref{fig:sk_velocity}a). This additional Magnus force contributes to the skyrmion motion in the x-axis, enhancing the skyrmion velocity. \par
To understand how the amplitude of standing and travelling SAWs that make up the orthogonal SAWs affect the skyrmion velocity, we simulated the skyrmion motion driven by orthogonal SAWs with different configurations. We fixed the standing SAW amplitude at 0.007 and varied the travelling SAW amplitude from 0.001 to 0.06. The skyrmion shows horizontal motion without transverse motion when the travelling SAW amplitude is less than 0.004. The skyrmion velocity increases as the travelling SAW amplitude increases from 0.001 to 0.004 (as shown in Fig.~\ref{fig:sk_velocity}c). We also fixed the travelling SAW amplitude at 0.003 and changed the standing SAW amplitude from 0.003 to 0.010. The skyrmion transverse motion can be sufficiently suppressed if the standing SAW amplitude is higher than 0.005. In other words, the strain amplitude of the standing SAWs needs to be higher than that of the travelling SAWs to provide enough pinning energy to confine the skyrmion motion in the $x$ direction. The skyrmion velocity gradually decreases with increasing standing SAW amplitude. \par

\begin{figure*}
\includegraphics[width=.6\textwidth]{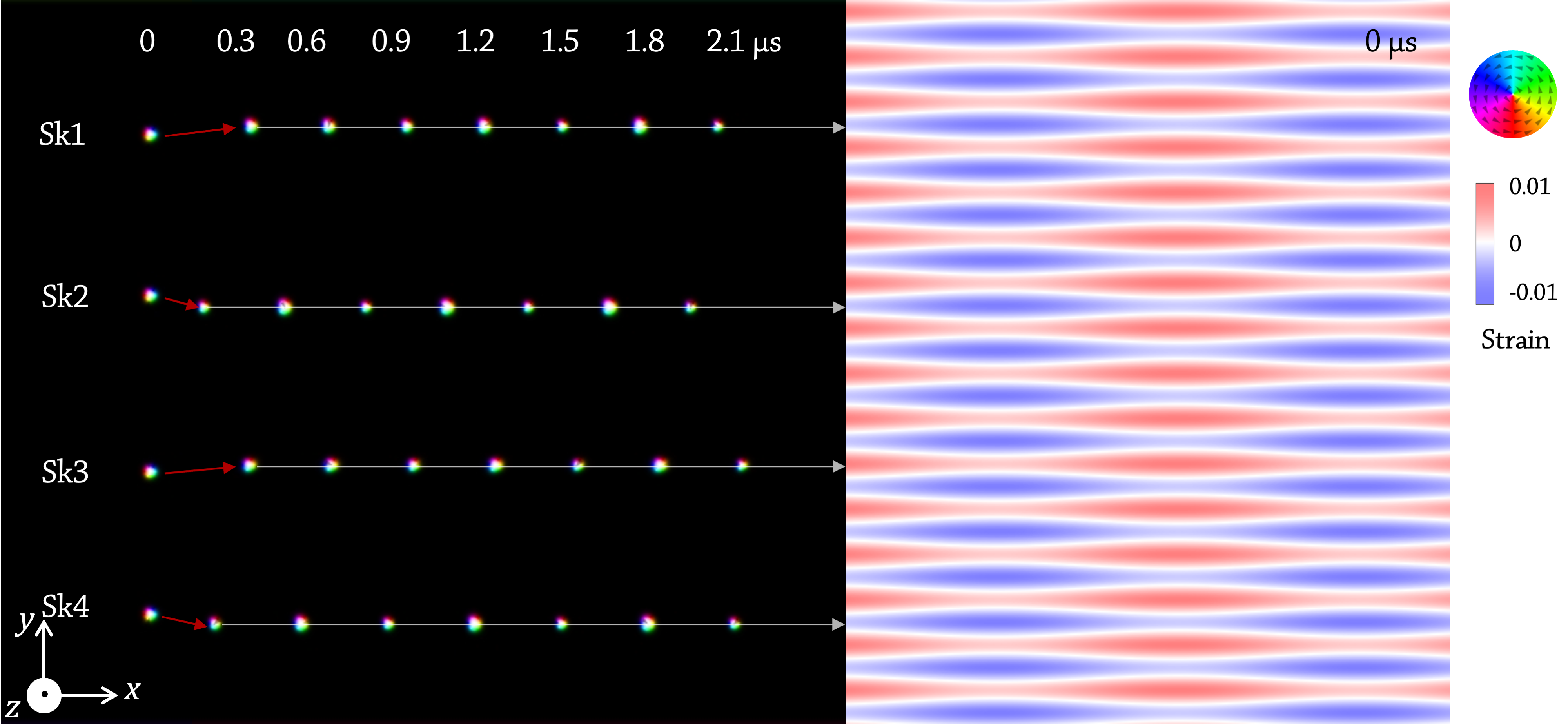}
\caption{\label{fig:ProposedDevice} 
Left-hand side of the graph: \revise{skyrmion (sk)} motion in the multichannel racetrack driven by orthogonal SAWs. Four skyrmions (labelled as Sk1 to Sk4) were initialised randomly along $y$ direction but the same $x$ position. Numbers on the top of the graph indicate the time duration (unit in \textmu s). Arrows indicate the skyrmion motion direction, which are in line with the anti-nodes of standing SAWs. Right-hand side of the graph: spatial strain profile at 0 \textmu s. \revise{Wavelengths and strain amplitudes of standing SAWs and travelling SAWs, which together form orthogonal SAWs, are 512 nm and 0.003, and 64 nm and 0.007, respectively. The maximum strain is 0.01.} \revise{The color code for the in-plane magnetisation component is shown by a color wheel. The Cartesian coordinates define the spatial orientation of the thin film.} }
\end{figure*}

Finally, we propose a multi-channel skyrmion racetrack obtained using orthogonal SAWs. In this simulation, we use the same orthogonal SAW property as above (i.e. the wavelengths and strain amplitudes of standing SAWs and travelling SAWs, which together form orthogonal SAWs, are 512 nm and 0.003, and 64 nm and 0.007, respectively) but with a larger space comprising $2048\times1024\times1$ nm\textsuperscript{3}. We initialised four skyrmions (Sk1 to Sk4 in Fig.~\ref{fig:ProposedDevice}) with random $y$ coordinates but with the same $x$ coordinates. As shown in Fig.~\ref{fig:ProposedDevice}, skyrmions are transported different distances during the first 0.3 \textmu s. This is owing to the fact that standing SAWs push skyrmions towards their nearest anti-nodes. This means that standing SAWs provide a driving force with the same/opposite direction as/to travelling SAWs depending their positions relative to the anti-nodes (see Fig.~\ref{fig:standingSAW}a). From 0.3 to 2.1 \textmu s, the velocity of all skyrmions remains the same. This is because the driving force is the same for all skyrmions once they arrive the pinning channel. With this design, one can create multiple channels to transport skyrmions without them interacting with each other, thereby increasing the amount of data that can be transported. The width and density of the channel can be determined by the wavelength of standing SAWs component of the orthogonal SAW.

\revise{It is important to note that the simulations in this study were conducted under ideal conditions to isolate and examine the effects of SAWs on skyrmion motion. This approach avoids the complexities introduced by defects, temperature variations, and other external factors. However, for practical device implementation, these factors must be considered. Theoretically, SAWs can drive skyrmion motion without introducing Joule heating, thereby offering a method that is attractive from an energy efficiency standpoint. However, in practical applications, heating effects are still observed when rf power is applied to interdigitated transducers~\cite{shuai2023separation}. A notable limitation of SAW-driven skyrmion motion is that the velocities achieved have not yet to match those in high current-driven skyrmion systems. Therefore, a direct comparison of the efficiencies of current-driven and SAW-driven skyrmion motion remains challenging. Nevertheless, a significant advantage of SAW is its capability to propagate over centimeter scales with minimal amplitude decay. This feature enables a single pair of transducers to simultaneously control thousands of nanostructures, which can significantly reduce the power consumption per nanostructure.}

% \section{Conclusion}
To conclude, we have demonstrated the skyrmion motion by SAWs with different modalities. Skyrmions were moved by standing SAWs and travelling SAWs in both longitudinal and transverse directions owing to the strain gradient that SAW introduced. Standing SAWs created pinning sites at their anti-nodes. In contrast, travelling SAWs provided a constant driving force to move skyrmions continuously. By combining longitudinal travelling SAWs and transverse standing SAWs, we have demonstrated the transport of skyrmions in the longitudinal direction without transverse motion by orthogonal SAWs. Our study suggests the possibility of multi-channel skyrmion racetrack memory/logic devices using SAWs.\par

% \section*{Conflict of Interest}
\vspace{0.5cm}
The authors have no conflicts to disclose.

% \section*{DATA AVAILABILITY}
\vspace{0.5cm}
The data that support the findings of this study are available in the University of Leeds repository, DOI:TBC.

% \begin{acknowledgments}
\vspace{0.5cm}
The authors gratefully acknowledge funding from the European Union’s Horizon 2020 research and innovation programme under the Marie Skłodowska-Curie grant agreement No. 860060 “Magnetism and the effect of Electric Field” (MagnEFi). 
J.S. expresses sincere gratitude for the invaluable contributions made by Dr Rutvij Bhavsar during the manuscript preparation process.
% \end{acknowledgments}

% \bibliographystyle{phys}
\bibliography{reference}% Produces the bibliography via BibTeX.

\end{document}